\documentclass[%
reprint,
superscriptaddress,
amsmath,
assume,
aps,
pra,
longbibliography
]{revtex4-1}
\usepackage{indentfirst}
\usepackage{graphicx}
\usepackage{dcolumn}
\usepackage{bm}
\usepackage[T1]{fontenc}
\usepackage{times}
\usepackage{color}
\usepackage[cmyk]{xcolor}
\usepackage{amsmath,amsfonts,amssymb,mathtools}
\usepackage{bbold}
\usepackage{tensor}
\usepackage{float}

\begin{document}

\preprint{APS/123-QED}

\title{ Long-range dipole-dipole exchange-induced atomic grating }%
\author{Xuan-Qian Bao}
 \affiliation{Center for Quantum Sciences and School of Physics, Northeast Normal University, Changchun 130024, P. R. China.}

\author{Xue-Dong Tian}%
\email{snowtxd@gxnu.edu.cn}
\affiliation{School of Physical Science and Technology, Guangxi Normal University, Guilin 541004, P. R. China.}

\author{Dong-Xiao Li}%
\affiliation{College of Physics Science and Technology, Shenyang Normal University, Shenyang 110034, P. R. China.}

\author{Yi-Mou Liu}%
 \email{liuym605@nenu.edu.cn}
 \affiliation{Center for Quantum Sciences and School of Physics, Northeast Normal University, Changchun 130024, P. R. China.}


\date{\today}

\begin{abstract}

We propose a theoretical scheme for dipole exchange-induced grating (DEIG) based on a hybrid system consisting of ultra-cold Rubidium ($^{87}$Rb) atomic ensemble and movable Rydberg spin atoms. The optical response of the grating appears as a superposition of three- and four-level configurations, similar to the cooperative optical nonlinear effect caused by the dipole blockade effect. However, such Rydberg atomic grating uniquely responds to the spatial positions of spin atoms, offering a novel approach to dynamically control electromagnetically induced gratings (EIG) except for input probe intensity. 
\end{abstract}

\maketitle


\section{Introduction}
The laser-induced coherence between atomic states modifies the optical response of atomic media.~Electromagnetically induced transparency (EIT) \cite{EIT1, EIT2, EIT3, EIT4}, well-developed theoretically and experimentally, serves as the foundation for numerous phenomena in nonlinear optics and quantum optical processes \cite{NQO1, NQO2, NQO3}. It also has a wide range of applications in slow or stopped light and optical storage \cite{SL1, SL2, SL3, SL4, SL5}, quantum information \cite{QI1, QI2, QI3}. 

 In fact, in some physical systems, coherent effects similar to EIT can also be achieved through various physical mechanisms.~For the case of an empty cavity, the driving laser can be substituted with a resonator's strongly coupled electromagnetic mode linked to the corresponding atomic transition \cite{VIT1, VIT2}, which leads to vacuum-induced transparency (VIT). In the optomechanics cavity system \cite{OCS1}, optomechanically induced transparency (OMIT) \cite{OMIT1, OMIT2} is another quantum effect caused by quantum interference between probe and anti-Stokes light induced by a mechanically resonant cavity.~Additionally, in interacting atomic systems, the similar transparency effects also include dipole-induced electromagnetic transparency (DIET) \cite{DIET1}, dipolar exchange-induced transparency (DEIT) with Rydberg atoms \cite{DEIT}.~Here, DEIT effect, proposed by D. Petrosyan, utilizes the strong dipole-dipole exchange interactions between atomic ensembles with one or more spin atoms to achieve interference in the transition pathways of weak light.
 
 As the combination of diffraction grating \cite{Grating1, Grating2, Grating3} and quantum coherence techniques, electromagnetically induced grating (EIG) can be obtained by applying standing-wave (SW) fields instead of traveling-wave (TW) fields in an EIT medium \cite{EIGs1, EIGs2, EIGs3, EIGs4}.~Such periodic structures can be used for all-optical control of light propagation \cite{AOC2}, two-photon spectral shaping \cite{TPSS}, atomic localization \cite{AL}, electromagnetically induced Talbot effect \cite{TE1, TE2}, and coherently induced photonic bandgap \cite{PB1, PB2, PB3}. Many schemes have been proposed \cite{HOD1, HOD2, HOD3, HOD4} to improve and modify the high-order diffraction efficiency in the EIG medium early on.
 
 In recent years, unconventional optical modulation has been introduced into the EIG structure to explore novel diffraction modes or new modulation methods. Based on increasingly attractive optical parity-time symmetric ($\mathcal{PT}$) \cite{PT1, PT2008-03, PT2013, PT2014-04, PT2014-06} and parity-time antisymmetric ($\mathcal{APT}$) optical systems \cite{APT2014, PT4, PT5}, asymmetric optical diffraction gratings provide a new diffraction mode \cite{PTgrating1, PTgrating2, PTgrating3, PTgrating4, PTgrating5, PTgrating6}. It is notable to achieve perfect positive or negative angle diffraction in symmetric structures. Additionally, utilizing the dipole-blockade effect of Rydberg states \cite{Rydberg1, Rydberg2, Rydberg2011, Rydberg2013}, schemes of cooperative nonlinear grating (CNG) \cite{cooperative2, Liu2016, cooperative3, cooperative4, cooperative6} have been proposed to explore the novel EIG's modulation approach or method. These types of gratings can distinguish light fields with different photon statistics and the diffraction characters of grating are sensitive to the input intensity of the quantum probe field.~The optical response of CNG resembles a combination of various level structures, which presents intriguing features absent in conventional EIGs.~Nevertheless, the challenges brought by nonlinear absorption, coupled with the confusion about utilizing the properties of probe light as a degree of control, have constrained its application.~Despite this, considering its unique properties, achieving similar optical control remains an important issue for exploration.

\begin{figure*}[ptb]
\includegraphics[width=1\textwidth]{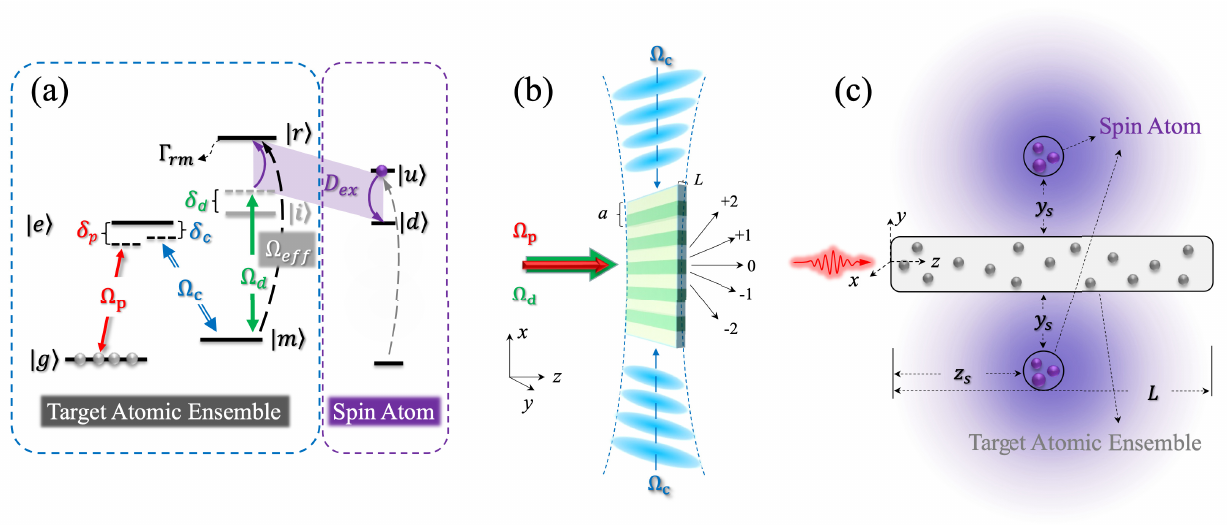}
\caption{(a) Schematic of a four-level $\mathcal{N}$ configuration in ultra-cold atom system. Three laser beams couple the system: the quantum probe field $\Omega_p$, the standing wave coupling field $\Omega_c$, and the traveling wave coupling field $\Omega_d$. The effective coupling field $\Omega_{\mathrm{eff}}$ coupled to the $|r\rangle$ state is mediated by the dipole-dipole exchange interaction $D_{ex}$ with the effective spin $J$ in panel(a). The spin atoms on each side with the Rydberg states $|u\rangle$ and $|d\rangle$, are confined to a small volume to form the effective spin $J = n_s/2$ which interacts with the medium atoms through the dipole-dipole exchange interaction. (b) Diffraction of a quantum probe field as it passes through a periodically modulated atomic medium. (c) The probe field $\Omega_p$ propagates along the z-axis in the optically dense atomic medium with linear density $\rho_0$ and length $L$. The dipole-dipole exchange interaction between the atom (at position $z$) and the effective spin $J$ (at position $r_s=\sqrt{z_s^2+y_s^2}$) leads to the DEIT of the probe field.}
\label{Fig1}
\end{figure*}

In this paper, we present a theoretical scheme of EIG featuring a five-level quasi-$\mathcal{N}$-type configuration structure within an ultra-cold ${ }^{87} \mathrm{Rb}$ atomic ensemble system, as depicted in Figure~\ref{Fig1}.~Positioned externally to the target ensemble, gate spin atoms with highly excited Rydberg states offer non-resonant long-range dipole-dipole exchange interactions with the atoms of the target ensemble. Distinguishing from the previous work \cite{Liu2016}, the new grating is based on the dipole exchange-induced transparency theory, introducing a new mechanism (the spin-atom position $z_s$) to modulate the diffraction characteristics. 

This work is organized through the following Sec.~\ref{SecII}, where we describe the background model based on dipole exchange-induced transparency effect, and Sec.~\ref{SecIII}, where we discuss the propagation characteristics of the weak field in this hybrid system and the modulation methods for the far-field diffraction properties of the new grating. We summarize, at last, our conclusions in Sec.~\ref{SecIV}.

\section{Model and Equations}\label{SecII}

We consider a hybrid system consisting of target atomic ensemble and the gate atoms (spin) situated outside the target ensemble, utilizing ultra-cold $^{87}$Rb atoms. The atoms of target ensemble have a five-level quasi-$\mathcal{N}$-type configuration, including the ground state $|g\rangle$, the excited state $|e\rangle$, the metastable state $|m\rangle$, and highly excited Rydberg state $|i\rangle$ and $|r\rangle$, as illustrated in Fig.~\ref{Fig1}(a).~A quantum field $\hat{\Omega}_{p}(z)=\hat{\mathcal{E}}_{p}(z)\cdot \wp_{g e} \sqrt{\omega_{p} /\left(2 \hbar \varepsilon_{0} V\right)}$, whose frequency is $\omega_{p}$, probes the transition $|g\rangle\leftrightarrow |e\rangle$ with detuning $\delta_{p}=\omega_{p}-\omega_{e g}$.~Here $\hat{\mathcal{E}}_{p}(z)$ ($\wp_{g e}$) is the local amplitude operator (transition dipole moment from $|g\rangle \leftrightarrow |e\rangle$) and $V$ is the local quantum volume.~The other two transitions $|e\rangle \leftrightarrow |m\rangle$ and $|m\rangle \leftrightarrow |i\rangle$ are coupled by classical optical fields $\Omega_c=\mathcal{E}_c\cdot\wp_{em}/2\hbar$ and $\Omega_d=\mathcal{E}_d\cdot\wp_{im}/2\hbar$, with detunings $\delta_{c}=\omega_{c}-\omega_{e m}$ and $\delta_{d}=\omega_{d}-\omega_{im}$, respectively. $\mathcal{E}_{c, d}$ represents the slow-varying amplitude of the coupling field with the transition dipole moment $\wp_{\alpha\beta}$ on transition $|\alpha\rangle\leftrightarrow |\beta\rangle$ ($\alpha,\beta\in\{e, m, i\}$).

The prepared spins represent the Rydberg states $|u\rangle$ and $|d\rangle$ of the gate atoms with transition frequency $\omega_{ud}$ and spin up (down) operator $\hat{\sigma}_{+}=|u\rangle\langle d|$ ($\hat{\sigma}_{-}=|d\rangle\langle u|$). Therefore, $\mathcal{D}_{ex}$ describes the spin-exchange interaction between the target atomic ensemble and the gate spin, with 
\begin{align}\label{Dex}
\mathcal{D}_{ex}=\frac{1}{4\pi\hbar\varepsilon_0}[\frac{\wp_{ri}\cdot\wp_{du}}{\mathbf{R}^3}-3\frac{(\wp_{ri}\cdot \mathbf{R})(\wp_{du}\cdot \mathbf{R})}{|\mathbf{R}|^5}].
\end{align}
Here $\wp_{ri}$ ($\wp_{du}$) is the dipole moment of the atomic transition between the Rydberg states $|r\rangle$ and $|i\rangle$ ($|d\rangle \leftrightarrow|u\rangle$).
The interaction can be easily attained as $\mathcal{D}_{ex}(z)=\frac{C_3}{|ze_z-r_s|^3}$ assuming $\wp_{ri}\parallel\wp_{du}\perp \mathbf{R}$ with $\mathbf{R}\equiv(z\hat{e}_z-\mathbf{r}_s)$ being the relative position vector between an atom (at $z$) and a spin (at $r_s$) where $C_3\equiv\frac{\wp_{ri}\cdot\wp_{du}}{4\pi\hbar\varepsilon_0}$. Symmetrically positioning the gate spins on both sides of the target atomic ensemble, the exchange interaction strength experienced by the target ensemble at the $z$-position is denoted as $\mathcal{D}_{ex}(z,  z_s)=\frac{2C_3}{\sqrt{(z-z_s)^2+y_s^2}^3}$, shown in Fig.~\ref{Fig1}(c).

Assuming the probe field $\hat{\mathcal{E}}_{p}(z,t)$ propagates along the $z$-axis, we adopt $\hat{P}\left(z,t\right) =\sqrt{\rho(z)}\hat{\tilde{\sigma}}_{ge}(z,t)$ to
describe the slowly-varying polarization, $\hat{S}_{m}\left(z,t\right)
=\sqrt{\rho(z)}\hat{\tilde{\sigma}}_{gm}(z,t)$ and $\hat{S}_{r}\left( z,t\right)  =\sqrt{\rho(z)}\hat{\tilde{\sigma}}_{gr}(z,t)$ to describe the spin fields with $\rho(z)$ being the atomic volume
density at position $z_j$. Here the average atomic transition operators are defined as $\hat{\tilde{\sigma}}_{ge}(z,t)=\frac{1}{\rho(z)}\sum^{\rho(z)}_j\hat{\sigma}_{ge}^je^{-ik_p z_j}$, $\hat{\tilde{\sigma}}_{gm}(z,t)=\frac{1}{\rho(z)}\sum^{\rho(z)}_j\hat{\sigma}_{gm}^je^{-i(k_p-k_c) z_j}$ and $\hat{\tilde{\sigma}}_{gr}(z,t)=\frac{1}{\rho(z)}\sum^{\rho(z)}_j\hat{\sigma}_{gr}^je^{-i(k_p+k_d)z_j}$ in the small volume $\Delta V$ centered at $z_j$, respectively, with $\hat{\sigma}^j_{\mu\nu}=|\mu_j\rangle
\left\langle \nu_j\right\vert $ representing single atomic transition operator for $j$th atom in the target ensemble ($\mu$, $\nu$ $\in\{g,e,m,r\}$ and $\mu\neq \nu$).

For the gate spin atoms, we choose $J=n_s/2$ and $J_{\pm}\equiv\sum_{j}^{n_s}\hat{\sigma}_{\pm}^{(s)}$ to represent collective spin and spin up (down) operators.~Effective coupling strength is given by $\tilde{\Omega}_{\mathrm{eff}}(z,z_s)=2n_s\Omega_{\mathrm{eff}}(z,z_s)=\frac{ 2n_s\Omega _d\cdot \mathcal{D}_{ex}(z)}{\delta _d}$, shown in Fig.~\ref{Fig1}(c), with state $|i\rangle$ adiabatically eliminated under the conditions of single photon off-resonant ($\delta_d\gg\Omega_d\sim\Gamma_{im}$) and two-photon resonance ($\delta_{d}+\delta_r=0$) [See Fig.~\ref{Fig1}(a)]. 

Then it is easy to write down the total interaction Hamiltonian $\hat{\mathcal{H}}=\hat{\mathcal{H}}_{p}+\hat{\mathcal{H}}_{a}+\hat{\mathcal{H}}_{ap}+\hat{\mathcal{U}}_{ex}+\hat{\mathcal{U}}_{vdW}$ to describe the quasi-four level system with the kinetic term $\hat{\mathcal{H}}_{p}$, an unperturbed atomic part $\hat{\mathcal{H}}_{a}$, an atomic-field interaction part $\hat{\mathcal{H}}_{af}$, the spin exchange interaction part $\hat{\mathcal{U}}_{ex}$, and the $vdW$ interaction term $\hat{\mathcal{U}}_{vdW}$, respectively:
\begin{widetext}
\begin{align}
     \hat{\mathcal{H}}_{p} & =\frac{i\hbar c}{L}\int_{0}^{L}dz\hat{\mathcal{E}}^{\dag}_{p}(z,t)\hat{\mathcal{E}}_{p}(z,t),\nonumber\\
     \hat{\mathcal{H}}_{a} & =-\hbar\int_{0}^{L}dz \left[\delta_{p}\hat{P}^{\dag}(z,t)\hat{P}(z,t)+\Delta_{1}\hat{S}^{\dag}_m(z,t)\hat{S}_m(z,t)+\Delta_{2}\hat{S}^{\dag}_r(z,t)\hat{S}_r(z,t)\right],\nonumber\\
     \hat{\mathcal{H}}_{ap} & = -\hbar\int_{0}^{L}dz\left[g\sqrt{\rho(z)}\hat{\mathcal{E}}^{\dag}_{p}(z,t)\hat{P}(z,t)+\Omega_c^{*}\hat{S}^{\dag}_m(z,t)\hat{P}(z,t)+h.c.\right],\nonumber\\
     \hat{\mathcal{U}}_{ex} & =-\hbar\int_{0}^{L}dz [\hat{S}^{\dag}_r(z,t)\hat{S}_m(z,t)\otimes\sum_{s}^{n_s}\Omega_{\mathrm{eff}}(z,z_s)\hat{\sigma}_{-}^{s}+h.c.],\nonumber\\
     \hat{\mathcal{U}}_{vdW} & =\frac{\hbar}{2}\int_{0}^{L}dz \int_{0}^{L}dz^{\prime}\hat{S}^{\dag}_r(z,t)\hat{S}^{\dag}_r(z^{\prime},t)\Delta(z-z^{\prime})\hat{S}_r(z^{\prime},t)\hat{S}_r(z,t),
\label{Eq_H1}
\end{align}
\end{widetext}
with $g=\wp_{ge}\sqrt{\omega_{p}/(2\hbar\epsilon_{0}V)}$ the
single-photon coupling strength and the multiple detunings $\Delta_1=\delta_p-\delta_c$ and $\Delta_2=\delta_p-\delta_c+\delta_d+\delta_r$. According to Eq.~$(\ref{Eq_H1})$, the dynamics of the system can be governed by Heisenberg-Langevin equations:
\begin{widetext}
\begin{align}\label{Eq_DME}
\partial_t\hat{\mathcal{E}}_{p}(z,t) & = -c \partial_{z} \hat{\mathcal{E}}_{p}(z,t)+\mathrm{i} g V\sqrt{\rho(z)}\hat{P}(z,t), \nonumber\\
\partial_t\hat{P}(z,t) & = [\mathrm{i}\delta_{p}-\gamma_{e}]\hat{P}(z,t)+\mathrm{i} \Omega_{c}^{*}\hat{S}_m(z,t)-\mathrm{i}g \hat{\mathcal{E}}^{\dag}_{p}(z,t)+\hat{F}_{ge},\nonumber \\
\partial_t\hat{S}_m(z,t) & = [\mathrm{i}\Delta_1-\gamma_{m}] \hat{S}_m(z,t)+\mathrm{i} \Omega_{c} \hat{P}(z,t)+\mathrm{i} \Omega_{\mathrm{eff}}(z,z_s) J_{-} \hat{S}_r(z,t)+\hat{F}_{gm},\nonumber\\
\partial_t\hat{S}_r(z,t) & = [\mathrm{i}(\Delta_2-\langle\hat{\Delta}_s\rangle)-\gamma_{r}] \hat{S}_r(z,t)+\mathrm{i} \Omega_{\mathrm{eff}}^{*} J_{+}\hat{S}_m(z,t)+\hat{F}_{gr},\\
,\nonumber
\end{align}
\end{widetext}
with the average associated Langevin noise operators $\langle \hat{F}_{g\mu} \rangle$ = 0 ($\mu=\{e,m,r\}$), where $\hat{\Delta}_{s}=\frac{1}{2}\int dz^{\prime}\hat{S}_{2}^{\dagger}(z^{^{\prime}},t)\Delta(z-z^{\prime})\hat{S}_{2}(z^{\prime},t)$ refers to the
$vdW$-induced frequency shift from $C_6(n_r)$,~making Eq.~\eqref{Eq_DME} difficult to solve.~To deal with
this dilemma, we consider all $n_{sa}=4\pi\rho(z) R_{b}^{3}/3$ atoms in the blockade sphere as a superatom (SA), with radius $R_{b}\approx\sqrt{C_6(n_r)\gamma_{e}%
/(\left\vert \Omega_{c}\right\vert ^{2}+\left\vert \Omega_{\mathrm{eff}}\right\vert^{2}\langle J_{+}J_{-}\rangle)}$. 

In the mean-field sense, $\langle\hat{\Delta}_s\rangle$ tends to infinite (vanishing) for an SA containing only one (yet no) Rydberg excitation, whereas $vdW$ interactions are regarded as absent between different SAs. Each SA is made up of four collective states $|G\rangle$, $|E^{(1)}\rangle$, $|M^{(1)}\rangle$, and $|R^{(1)}\rangle$ in the weak probe limit \cite{David2011}. Such an SA has transition operators defined by $\hat{\Sigma}_{GE}=|G\rangle\langle E^{(1)}|$, $\hat{\Sigma}_{GM}=|G\rangle\langle M^{(1)}|$, and $\hat{\Sigma}_{GR}=|G\rangle\langle R^{(1)}|$, whose dynamic evolutions obey equations similar to those for $\hat{P}(z)$, $\hat{S}_m(z)$, and $\hat{S}_r(z)$ with $g\hat{\mathcal{E}}_p$ replaced by $\sqrt{n_{sa}}g\hat{\mathcal{E}}_p$. In the weak probe limit, considering $\hat{\Sigma}_{RR}\simeq\hat{\Sigma}_{RG}\hat{\Sigma}_{GR}$ with $\hat{\Sigma}_{GG}\simeq\hat{\Sigma}_{RR}\simeq 1$, the Rydberg excitation of superatom $\hat{\Sigma}_{RR}$ (merely contributed by state $|\hat{R}^{(1)}\rangle$) can be written as 

\begin{align}\label{rho_rr}
\hat{\Sigma}_{RR} & \simeq \hat{\Sigma}_{RR}^{(1)}=|\hat{R}^{(1)}\rangle\langle\hat{R}^{(1)}|\nonumber\\
& = \frac{n_{sa}g^2\hat{\mathcal{E}}^{\dag}_{p}(z)\hat{\mathcal{E}}_{p}(z) \Omega_c^{*}\Omega_c|\Omega_{\mathrm{eff}}|^2\langle J_{+}J_{-}\rangle}{n_{sa}g^2\hat{\mathcal{E}}^{\dag}_{p}(z)\hat{\mathcal{E}}_{p}(z) \Omega_c^{*}\Omega_c|\Omega_{\mathrm{eff}}|^2\langle J_{+}J_{-}\rangle+DD},
\end{align}
where $DD=\Omega_c^4[\Delta_2^2+\gamma_r^2]+2\Omega_c^2[\Delta_2^2+\gamma_r^2][\gamma_m\gamma_e-\delta_p\Delta_1]+2\Omega_c^2\Omega^2_{\mathrm{eff}}\langle J_{+}J_{-}\rangle[\gamma_r\gamma_e+\delta_p\Delta_2]+(\delta_p^2+\gamma_e^2)\{[\Delta_1^2+\gamma_m^2][\Delta_2^2+\gamma_r^2]+\Omega_{\mathrm{eff}}^4\langle J_{+}J_{-}\rangle^2\}$. However, in the general case, other higher-order collective states ($|\hat{E}^{(1)}\hat{R}^{(1)}\rangle$, $|\hat{M}^{(1)}\hat{R}^{(1)}\rangle$, $|\hat{E}^{(1)}\hat{M}^{(1)}\hat{R}^{(1)}\rangle$, ...) are also essential to evaluate Rydberg excitation $\hat{\Sigma}_{RR}\simeq \hat{\Sigma}_{RR}^{(1)}+\hat{\Sigma}_{RR}^{(2)}+\hat{\Sigma}_{RR}^{(3)}+...$ [See \eqref{A2}, \eqref{A3} and \cite{Liu2014, Tian_2023}].
Conditioned upon $\langle\hat{\Sigma}_{RR}\rangle \to 1.0$
($\langle\hat{\Delta}_s\rangle\to\infty$) or $\to 0.0$ ($\langle\hat{\Delta}_s\rangle\to 0.0$), the $n_{sa}$ cold atoms in an SA behave either like a three-level $\Lambda$-type system or a four-level $\mathcal{N}$-type system. Substituting it in the field propagation equation \eqref{Eq_DME} without the time-derivative, we obtain the conditional probe polarizability:
\begin{align}\label{Polar}
\hat{P}(z)  &  =\hat{P}^{\left(\Lambda\right)  }(z)\langle\hat{\Sigma}%
_{RR}(z)\rangle  +\hat{P}^{\left(N\right)}(z)[1-\langle\hat{\Sigma}_{RR}%
(z)\rangle],\nonumber\\
\hat{P}^{\left(\Lambda\right)}(z)  &  =\frac{\mathrm{i}g\hat{\mathcal{E}}^{\dag}_{p}(z)[\gamma_{m}-\mathrm{i}\Delta_1]}{[\gamma_{e}-\mathrm{i}\delta_{p}][\gamma_{m}-\mathrm{i}\Delta_1]+\Omega_c^{*}\Omega_c},\\
\hat{P}^{\left(N\right) }(z)  &  =\frac{-\mathrm{i}g\hat{\mathcal{E}}^{\dag}_{p}(z)[\gamma_{m}^{\prime}\gamma_{r}^{\prime}+|\Omega_{\mathrm{eff}}(z,z_s)|^2\langle J_{+}J_{-}\rangle]}{\gamma_{e}^{\prime}\gamma_{m}^{\prime}\gamma_{r}^{\prime}+\gamma_{e}^{\prime}|\Omega_{\mathrm{eff}}(z,z_s)|^2 \langle J_{+}J_{-}\rangle+\gamma_{r}^{\prime}\Omega_c^{*}\Omega_c},\nonumber
\end{align}
 with effective decay $\gamma_{e}^{\prime}=\gamma_{e}-\mathrm{i}\delta_{p}$, $\gamma_{m}^{\prime}=\gamma_{m}-\mathrm{i}\Delta_{1}$, $\gamma_{r}^{\prime}=\gamma_{r}-\mathrm{i}(\Delta_2-\langle\hat{\Delta}_s\rangle)$. 
 To examine the probe response of the considered medium, we should further deal with the propagation equations of intensity $\mathcal{I}_p(z)=\langle \hat{\mathcal{E}}^{\dag}_p(z)\hat{\mathcal{E}}_p(z)\rangle\propto\langle \hat{\Omega}^{\dag}_p(z)\hat{\Omega}_p(z)\rangle$ and phase $\phi_p(z)=\arg\langle \hat{\mathcal{E}}^{\dag}_p(z)\rangle$. Starting from the first row of Eq.~\eqref{Eq_DME}, we attain 
\begin{align}\label{Ip and phi}
\partial_z\mathcal{I}_p(z) & =-\kappa(z)\langle \hat{\mathcal{E}}_p^{\dag}(z) \textbf{Im}[ \hat{\alpha}(z)]
\hat{\mathcal{E}}_p(z)\rangle,\nonumber\\
\partial_z\phi_p(z) & =\frac{\kappa(z)}{2}\left[\langle \hat{\alpha}(z)\rangle-\mathrm{i}\frac{\langle \hat{\mathcal{E}}_p^{\dag}(z)\textbf{Im}[\hat{\alpha}(z)]\hat{\mathcal{E}}_p(z)\rangle}{\mathcal{I}_p(z)}\right],
\end{align} 
with $\kappa(z)=\sqrt{\rho(z)}\omega_p|\wp_{eg}|^2/(\varepsilon_0\hbar c)$ denoting the resonant absorption coefficient and the local normalized probe susceptibility $\langle\hat{\alpha}(z)\rangle=\langle\hat{P}(z)/\sqrt{\rho(z)}g\hat{\mathcal{E}}_p(z)\rangle$, respectively, where susceptibility reads $\chi_p(z)=\rho(z)\wp_{ge}\langle\hat{\alpha}(z)\rangle/\hbar\varepsilon_0$.~Here we introduce the two-photon correlation $g_{p}^{(2)}(z,z^{\prime})=\frac
{\langle\mathcal{\hat{E}}_{p}^{\mathcal{\dag}}(z)\mathcal{\hat{E}}%
_{p}^{\mathcal{\dag}}(z^{\prime})\mathcal{\hat{E}}_{p}(z^{\prime})\mathcal{\hat{E}}%
_{p}(z)\rangle}{\langle\mathcal{\hat{E}}_{p}^{\mathcal{\dag}}%
(z)\mathcal{\hat{E}}_{p}(z)\rangle\langle\mathcal{\hat{E}}_{p}%
^{\mathcal{\dag}}(z^{\prime})\mathcal{\hat{E}}_{p}(z^{\prime})\rangle}$ to quantify the modification of photon statistics based on the dipole-blockade effect. With the above considerations, it is straightforward to expand Eq.~\eqref{Ip and phi} as
\begin{align}\label{Ip and phi 2}
\partial_z \mathcal{I}_p(z) & =-\kappa(z)\langle \textbf{Im}[\hat{\alpha}(z)]\rangle \mathcal{I}_p(z),\nonumber\\
\partial_z\phi_p(z) & =+\kappa(z)\langle \textbf{Re}[\hat{\alpha}(z)]\rangle /2,\\
\partial_z g_p^{(2)}(z) & =-\kappa(z)\{\langle\hat{\Sigma}_{RR}(z)\rangle \textbf{Im}[\hat{\alpha}^{\Lambda}(z)-\hat{\alpha}^{N}(z)]\rangle\} g_p^{(2)}(z),\nonumber
\end{align}
which can be numerically solved using the inatial conditions ($\mathcal{I}_p(0)$~and $g_p^{(2)}(0)=1$)~with Eqs.~\eqref{Ip and phi} and \eqref{Ip and phi 2}.

To investigate the properties of EIG, we replace the coupling field with a standing wave field (SW), whose Rabi frequency can be written as $\Omega_{c}(x)=\Omega_{c0}\cdot\sin^2\left[\frac{\pi (x-x_0)}{a}\right]
$. Here $a=\lambda_{c}/\sin\varphi$ is SW spatial period, and $\varphi$ denotes the angle between the control field $\Omega_{c}$ and the $z$-axis, with $x_0$ marking the intersection point of the optical axis with the $x$-axis. Under this condition, for the medium with a thickness of $L$ along the $z$ direction, we can obtain the transmission function:
\begin{align}
T_{L}(x) & = \mathcal{A}(x)\cdot e^{\mathrm{i}\Phi(x)},\\
\mathcal{A}(x) & =\langle\hat{\mathcal{E}}_p(x,L)\rangle/\langle\hat{\mathcal{E}}_p(x,0)\rangle,\nonumber\\
\Phi(x) & =\phi_p(x,L)-\phi_p(x,0),\nonumber
\label{TL}
\end{align}
where $\mathcal{A}(x)$ and $\Phi(x)$ represent the amplitude and phase components, respectively. The Fraunhofer diffraction equation or far-field intensity diffraction equation can be obtained from the Fourier transform of $T_{L}(x)$,
\begin{align}
\Omega^{I}(\theta) & =\int_{-a / 2}^{+a / 2} T_{L}(x) e^{-i 2 \pi x\cdot R \sin (\theta)} \mathrm{d}x,
\end{align}
and multi-beam interference
\begin{align}
I_{p}(\theta) & =\frac{\left|\Omega^{I}(\theta)\right|^{2} \sin ^{2}[M \pi R \sin (\theta)]}{M^{2} \sin ^{2}[\pi R \sin (\theta)]},
\end{align}
with $R=a/\lambda_p$. In addition, $\theta$ represents the diffraction angle of the probe photon relative to the $z$ direction, and $M$ represents the ratio of the beam width $w$ to the SW period $a$ $(M=w / a)$. The angle of the $\xi$th-order diffraction probe field is determined by $\xi=R \sin (\theta) \in(0,\pm 1,\pm 2, \cdots)$.
     
\section{Results and Discussion}\label{SecIII}

In this section, we examine the atomic ensemble's steady-state optical response and numerically verify the atomic grating's far-field diffraction properties. Firstly, we specify realistic parameters for the ultra-cold ${}^{87} \mathrm{Rb}$, specifically with the states as
$|g\rangle\equiv5S_{1/2}|F=2\rangle$, $|e\rangle\equiv5P_{1/2}|F=2\rangle$, $|m\rangle \equiv5S_{1 / 2}|F=1\rangle$, $|i\rangle \equiv 64 P_{1 / 2}$, $|r\rangle \equiv 65 S_{1/ 2}$, $|u\rangle \equiv 60 P_{3 / 2}$, $|d\rangle \equiv 60 S_{1 / 2}$.~The corresponding decay rates are $\gamma_{g e}=2 \pi\times5.9$ MHz, $\gamma_{g r}=2.15$ kHz, $\gamma_{g m}\approx1$ kHz, respectively, with $C_3\equiv \wp_{ri}\cdot\wp_{du}/(4\pi\epsilon _0\hbar) = 2\pi\times$ 2.281 GHz$\cdot\mu m^3$ \cite{ARC1, ARC2}. Additionally, we assume that the target atomic ensemble has a length $L$ = 250 $\mu$m along $z$-axis with atomic density $\rho_0=8.0\times 10^{10}$ cm$^{-3}$.~Considering the quasi-one-dimensional assumption for the target atomic ensemble, the spin exchange interaction reads $\mathcal{D}_{ex}(z_s,z)=C_3/[(z_s-z)^2+\overline{y}_s^2]^{3/2}$, with $\Delta y\ll y_s$ ($\overline{y}_s\simeq y_s+\Delta y$).~Here $\overline{y}_s$ represents the average distance between the ensemble atoms and the spin atoms, with $y_s$ ($\Delta y$) being the distance between spin (ensemble) atoms and the boundary of the ensemble.~Under these conditions, the effective coupling, denoted as $\tilde{\Omega}_{\mathrm{eff}}(z_s,z)=2n_s\Omega_d \cdot \mathcal{D}_{ex}(z_s,z)/\delta_d$, is determined by $\delta z=|z_s-z|$, fixed $y_s=40\mu m$ and $\delta_d=20\times 2\pi$ MHz $=10\Omega_d$. In the appendix \ref{Omega}, we discuss the feasibility of taking a one-dimensional approximation.

\subsection{Susceptibility under the weak probing limit}\label{Weak}

\begin{figure}
\includegraphics[width=0.48\textwidth]{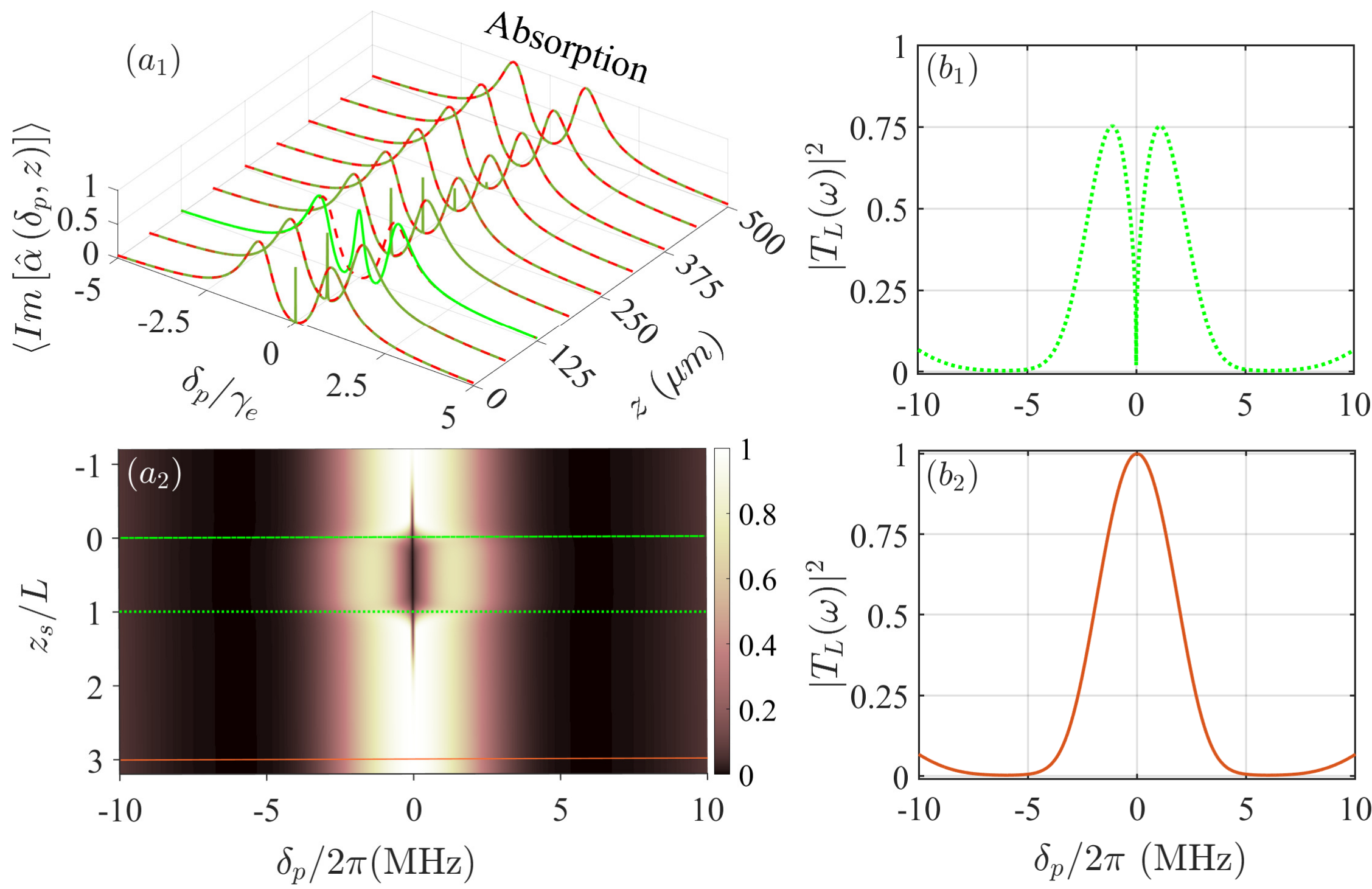}
\caption{Absorption $\langle\textbf{Im}[\hat{\alpha}]\rangle$ as a function of $\delta_p$ and $z$ in the target ensemble with the gate spin located at $z_s=L/2$, where $n_s=\sqrt{\langle\hat{J}_{+}\hat{J}_{-}\rangle}=$ 500 (olive- or blue-solid curves), $n_s=0$ (red-dashed and dotted curves) in panel ($a_1$). The intensity transmissivity $|T_{L}|^2=\mathcal{I}_p(L)/\mathcal{I}_p(0)$ varying with probe detuning $\delta_p/2\pi$ and the spin position $z_s$ in panel ($a_2$). Panel ($b_1$) and ($b_2$) correspond green-dotted curves ($z_s=0$ or $L$) and orange-solid one ($z_s=3L$) of panel ($a_2$), respectively. We set the atomic ensemble with density being $\rho_0=8.0 \times 10 ^ {10}$ cm$^{-3}$, length $L=250\mu$m along $z$-axis and other parameters are chosen as $\gamma_{g e}=5.9\times2 \pi $ MHz, $\Omega_{p}= 0.001 \times 2 \pi$ MHz, $\Omega_{c}= 6.0 \times 2 \pi$ MHz, $\Omega_{d}=2.0 \times 2 \pi$ MHz, $\delta_{d}=10 \Omega_{d}$, and $y_{s}=40$ $\mu$m.}
\label{Fig2}
\end{figure}

We start by analyzing the steady-state optical response of a quasi-one-dimensional atomic ensemble with two traveling wave (TW) control fields ($\Omega_{c}=\gamma_{ge} \simeq 6.0 \times 2 \pi$ MHz, $\Omega_{d}=2.0 \times 2 \pi$ MHz). Under the weak probe limit ($\Omega_p/2\pi=0.001$ MHz), the excitation probability for Rydberg state $|r\rangle$ within the target atomic ensemble is sufficiently low to render the $vdW$ interactions between these $|r\rangle$ states negligible ($\hat{\mathcal{U}}_{vdW}\to 0$).~In this scenario, Fig.~\ref{Fig2}($a_1$) depicts the absorption spectra ($\langle\textbf{Im}[\hat{\alpha}]\rangle$) for atoms located at different position $z$ (along the $z$-axis), with the gate spin atoms at $z_s=L/2$.~In the absence of gate atomic spin excitation ($n_s=0$), the absorption spectrum follows a typical three-level $\Lambda$-type EIT curve, indicated by the red-dashed and dotted lines. With spin atoms prepared in the spin-up state $|u\rangle$ (spin exaction number $n_s=\sqrt{\langle\hat{J}_{+}\hat{J}_{-}\rangle}=$ 500 \cite{DEIT}), the optical response of the target atomic ensemble exhibits $\mathcal{N}$-type dual EIT spectra (olive curves) in Fig.~\ref{Fig2}($a_1$).~The effective coupling strength $\Omega_{\mathrm{eff}}(z,z_s)$ decreases with the increase of $z$ ($z>L/2$, olive curves), causing dual EIT transmission peaks to split and eventually merge into a single EIT peak ($\Lambda$-type, green curves) as $z$ increases.~This trend causes the position-dependent absorption $\textbf{Im}\langle\hat{\alpha}_p(\delta_p,z)\rangle$.

Figure~\ref{Fig2}(a$_2$) illustrates the variation of the intensity transmission spectrum of the target atomic ensemble $|T_{L}(\omega,z_s)|^2=\mathcal{I}_p(z=L)/\mathcal{I}_p(z=0)$ versus the gate spin position $z_s$. Here, the optical depth (OD) of the target ensemble is $\zeta =\frac{8\pi\rho_0\wp_{eg}^2L}{\varepsilon\hbar\lambda_p\Gamma_{eg}}\simeq 11.7$ with $L=250\mu$m and $\rho_0=8.0\times 10^{10}$cm$^{-3}$. When the spin atoms are near the target ensemble ($z_s\in [0, L]$), we will attain a dual-EIT-peak transmission spectrum of a typical four-level $\mathcal{N}$-type system, displayed by the range between green-dotted curves with spin positions $z_s=0$ or $z_s=L$ in Fig.~\ref{Fig2}(b$_1$).~In contrast, moving away the spin atoms from the target atomic medium ($z_s>L$), the normalized transmission spectrum converges to the typical $\Lambda$-type single-peak transmission spectrum, which is depicted by the orange-solid curves in Fig.~\ref{Fig2}(a$_2$) and Fig.~\ref{Fig2}(b$_2$) for the position $z_s=750\mu m=3L$. This intensity profile resembles the superposition of two distinct level structures arising from cooperative optical nonlinearity \cite{CON1, CON2, CON3, CON4}.

\begin{figure}
     \includegraphics[width=0.48\textwidth]{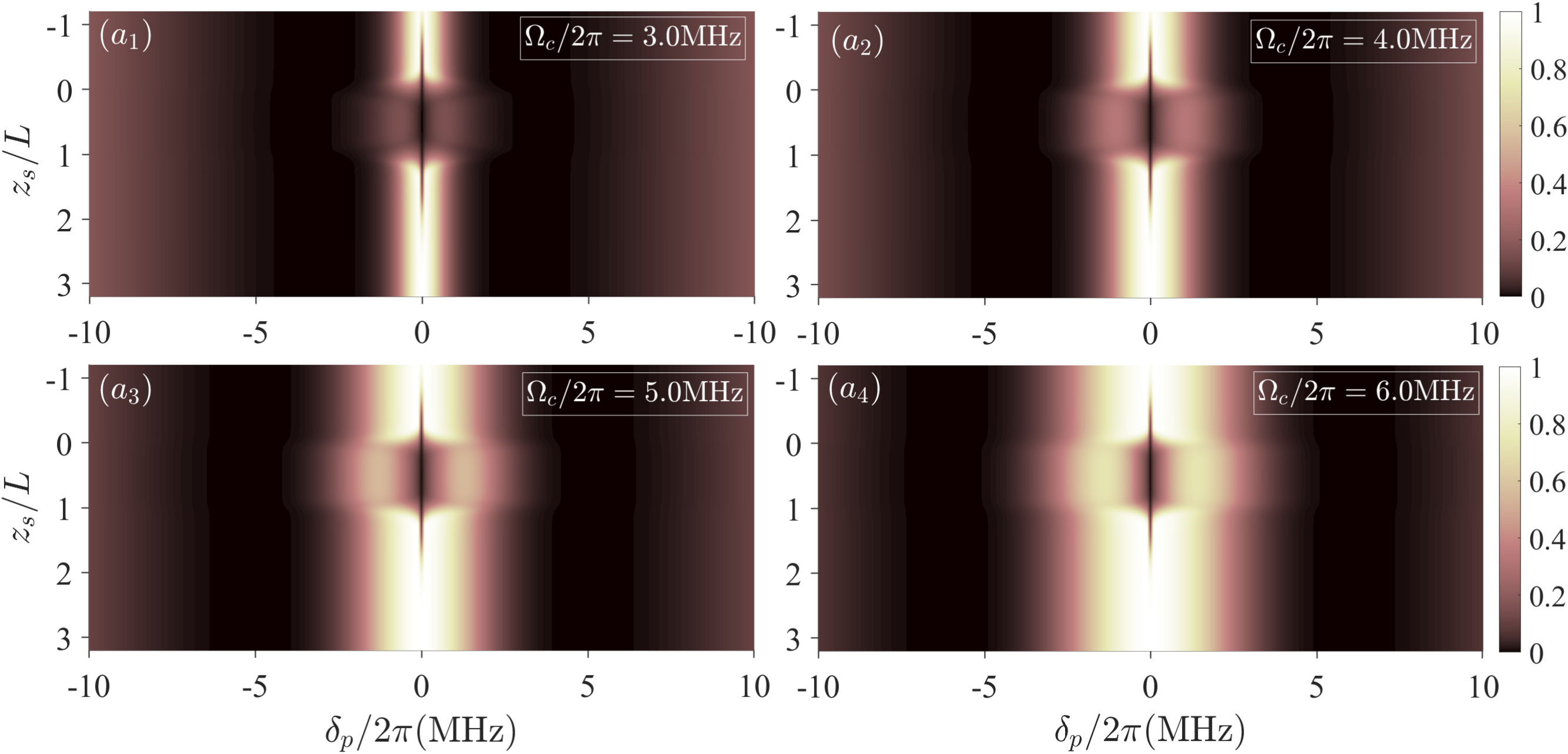}
     \caption{The intensity transmissivity $|T_{L}(\omega)|^2$ varying with probe detuning $\delta_p/2\pi$ and the spin position $z_s$, under different coupling strength $\Omega_c/2\pi=$ 3.0, 4.0, 5.0 and 6.0 MHz for panels (a$_1$) to (a$_4$), respectively.~The other parameters are the same as Figure~\ref{Fig2}.}
     \label{Fig3}
\end{figure}

Compared to Fig.~\ref{Fig2}(b$_2$) with $z_s =3 L$, imperfect transmission peaks ($|T(\omega)|^2\simeq 0.75$ in Fig.~\ref{Fig2}(b$_1$)) can be observed under the condition $z_s\in [0, L]$, attributed to the cooperative absorption effect. That means the ensemble atoms absorbe photons as $\Lambda$ type structures at the entrance ($z\sim 0$) and as $\mathcal{N}$-type structures in the middle ($z\sim L/2$). Consequently, the total transmission primarily influenced by the frequency overlap between the absorption peaks of the $\Lambda$- and $\mathcal{N}$-type structure. Photons, whose frequencies near resonance with the dual EIT peaks of $\mathcal{N}$-type structure (at $\delta_p\simeq \pm |\tilde{\Omega}_{\mathrm{eff}}(z_s)|$ corresponding to the \textit{dressed-states} $\left|\pm M^{(1)}\right\rangle=(|M^{(1)}\rangle\pm|R^{(1)}\rangle)/\sqrt{2}$), are strongly absorbed by a large number of atoms with $\Lambda$-type structure, particularly upon their initial propagation into the target ensemble.

It is also supported by Figure~\ref{Fig3}: the FWHM (full width at half maximum) of the single EIT spectrum for three-level $\Lambda$-type atoms $w_{3\Lambda}\propto\Omega_c^2/\sqrt{\gamma_e^2+\delta_p^2}$ expands as the coupling strength increases ($\Omega_c/2\pi=3.0, 4.0, 5.0$ and $6.0$ MHz), corresponding to Fig.~\ref{Fig3}(a$_1$)-(a$_4$).~Therefore, the transmission peaks of the dual EIT spectrum will have more overlap with the single transmission peak of the three-level system, and even fall completely within its spectrum in some cases, resulting in more photons passing through the atomic ensemble, performing the stronger dual EIT transmission peaks with $0<z_s<L$.

\subsection{Modulation for far-field diffraction} 

\begin{figure}
\includegraphics[width=0.48\textwidth]{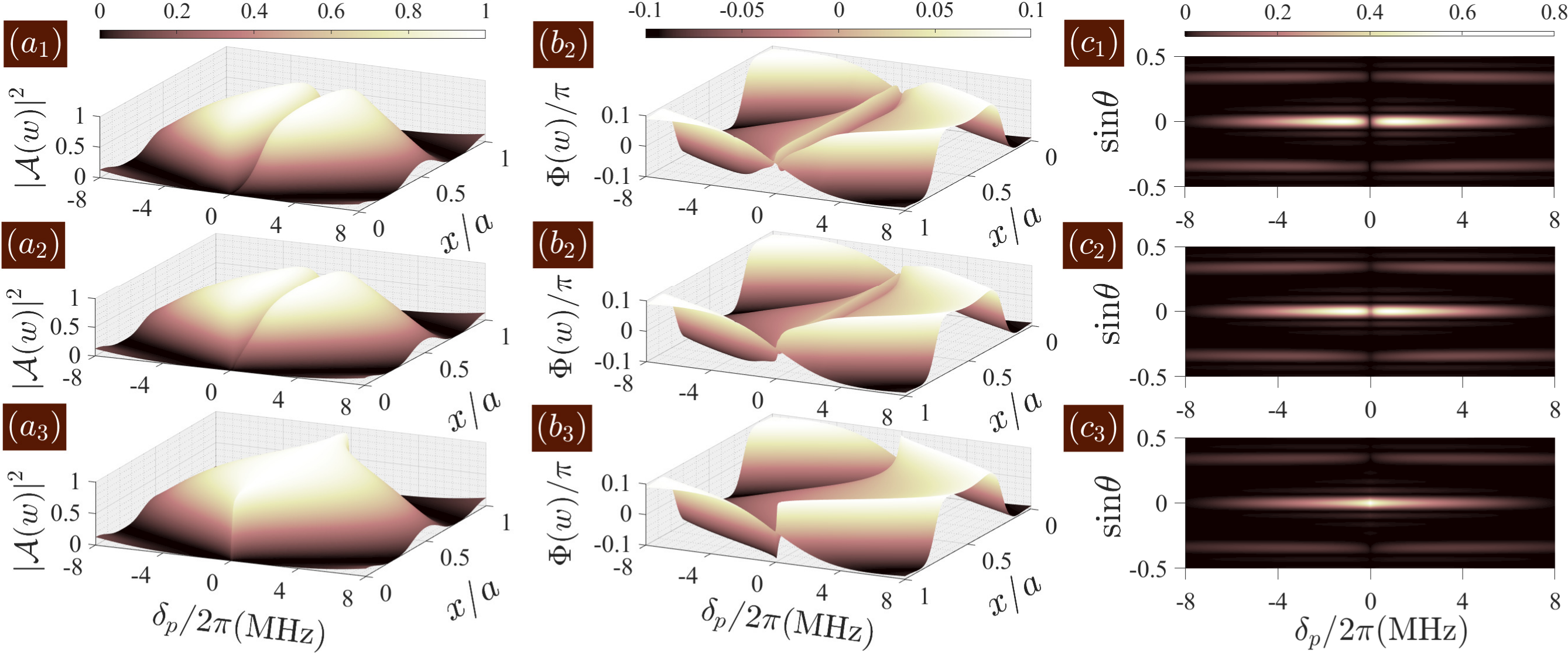}
\caption{Output amplitude transmission $|\mathcal{A}(\omega)|^2$ and phase $\Phi(\omega)$ versus detuning $\delta_{p} / 2 \pi$ and position $x/a$ in single period are shown in panel (a$_1$-a$_3$) and (b$_1$-b$_3$), respectively.~(c$_1$-c$_3$) Fraunhofer diffraction intensity $I_{p}(\theta)$ as a function of $\delta_{p} / 2 \pi$ and $\sin\theta$ for different spin locations $z_s$.~Panels from top to bottom correspond to the cases for gate spin positions $z_s=L/2$, $L$, and $3L$, with $\Omega_{c0}=2\gamma_{ge}\simeq 12\times 2\pi$ MHz, $R=3$, $M=5$, and other parameters are the same as Figure~\ref{Fig2}.}
\label{Fig4}
\end{figure}

Next, we will replace the coupling field with a standing wave ($\Omega_c=\Omega_{c_0}\cdot\sin\left[\pi (x-x_0)/a\right]$) and further discuss the diffraction properties of atomic gratings based on our system.
Figure~\ref{Fig4}(a$_1$)-(b$_3$) display the transmission information of the weak probe field including amplitude part $|\mathcal{A}(\omega)|^2$ and phase $\Phi(\omega)$ versus probe detuning $\delta_p/2\pi$ and position $x$ within one period ($x\in [0, a]$).~Locating the spin atoms progressively farther from $z_s=L/2$, a transition trend is observed in the target atomic medium's steady-state optical response, evolving from a three-level system's ($\Lambda$-type) to a four-level system's ($\mathcal{N}$-type) characteristics, which stems from the diminished dipole-dipole exchange interactions ($\mathcal{D}_{ex}\propto z_s^{-\frac{3}{2}}$ for $z_s>L/2$).~Substantiating this trend, Fig.~\ref{Fig4}(c$_1$-c$_3$) displays Fraunhofer diffraction intensity $I_p(\theta)$ versus diffraction angle $\theta$ and detuning of probe field $\delta_p$ with three different spin atomic positions ($z_s=L/2, L, 3L$). The observed diffraction pattern reveals a blend of $\mathcal{N}$-type pure-phase grating and $\Lambda$-type composite modulation grating, reminiscent of a cooperative nonlinear grating \cite{Liu2016}. However, in our scheme, the superimposed weights of different structures are determined by the spin position $z_s$, which provides a new degree of freedom for grating modulation.

\begin{figure}[t]
\includegraphics[width=0.48\textwidth]{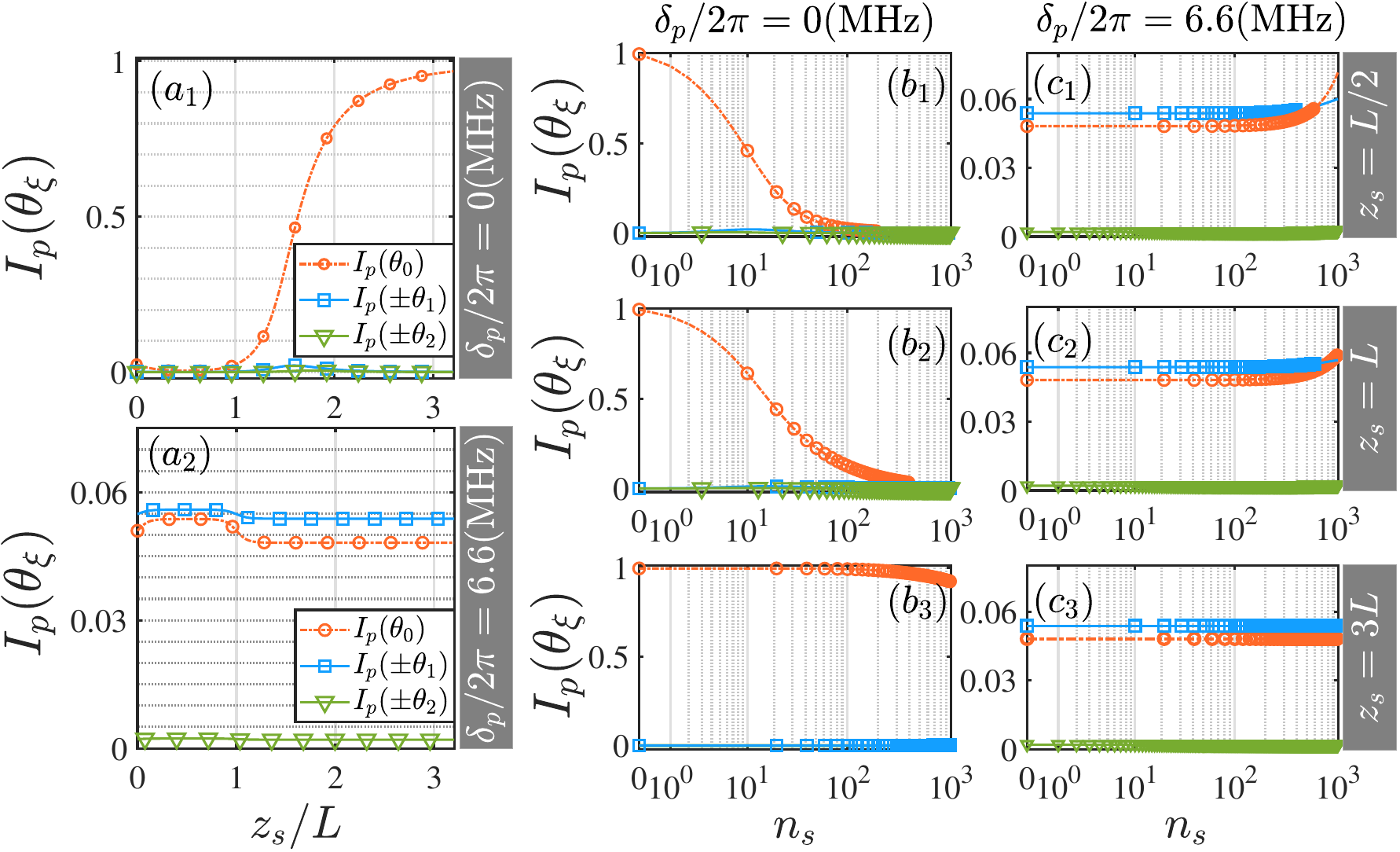}
\caption{The Fraunhofer diffraction intensity of the main order $I_p(\theta_{\xi})$ ($\xi\in{0,\pm1,\pm2}$, represented by orange circles, blue squares, and olive triangles) is plotted against $z_s$ for $\delta_p/2\pi=0$, $6.6$ MHz, as shown in panels (a$_1$) and (a$_2$), with the gate atoms' spin excitation number set at $n_s=500$. Panels (b$_1)$-(b$_3)$ and (c$_1)$-(c$_3)$ illustrate cases for $\delta_{p}/2\pi=0$, $6.6$ MHz, respectively. Sequentially, from top to bottom, the panels represent different spin positions $z_s=L/2$, $L$, and $3L$. Other parameters are the same as those in Figure~\ref{Fig4}.}
\label{Fig5}
\end{figure}

Figure~\ref{Fig5} (a$_1$) and (a$_2$) illustrate how the diffraction intensity of the principal orders $I_p(\theta_{\xi})$ ($\xi\in{0,\pm 1, \pm 2}$) varies with the spin position $z_s$, corresponding to different probe frequencies ($\delta_p/2\pi=0$ or $6.6$ MHz). The absorption of the target ensemble governs the diffraction intensity variation at the single EIT peak ($\delta_p=0$).~Near the region $0<z_s<L$, the absorption characteristics of the $\mathcal{N}$type system are dominant, resulting in a large amount of zero-order diffraction intensity $I_p(\theta_0)$ absorption at a single EIT peak. Simultaneously, the medium's weak dispersion is characterized by minimal total phase $\Phi(\omega)$ (in Fig.~\ref{Fig4}) and a near-zero real part of the susceptibility $\frac{2\omega}{c}\chi_p^{\prime}\simeq 0$, causing negligible high-order diffraction ($I_p(\theta _{\pm 1})\simeq I_p(\theta_{\pm 2})\to 0$). When detuning satisfies $\delta_p>\Omega_{c_0}/2$, the system attains considerable dispersion, due to the accumulation of sufficient phase supported by Fig.~\ref{Fig4}.~As shown in Fig.~\ref{Fig5}(a$_2$), the intensity of the first-order diffraction exceeds that of the zeroth-order, $I_p(\theta _{\pm 1})>I_p(\theta_0)$, which is also tunable by the spin position $z_s$.

On the other hand, spin excitation number $n_s$ also dominates the effective coupling of the transition $|m \rangle\leftrightarrow |r\rangle$, $\tilde{\Omega}_{\mathrm{eff}}(x,z_s)=2n_s\cdot\Omega_{\mathrm{eff}}(x,z_s)=2n_s\Omega_d D_{ex}(x,z_s)/\delta_d$, which modulates the splitting distance between dual EIT peaks $w_{d-\mathrm{EIT}}=2\tilde{\Omega}_{\mathrm{eff}}$ and the dispersion within this frequency bandwidth. Figure~\ref{Fig5}(b$_1$)-(c$_3$) depict the impact of the excitation number of the gate atomic spins ($n_s$) on the diffraction intensity of principal orders $I_p(\theta_{\xi})$ ($\xi\in\{0,\pm 1, \pm 2\}$) with $z_s=L/2, L, 3L$.~Focusing on the same probe detuning as those in Fig.~\ref{Fig5}(a$_1$) and (a$_2$) ($\delta_p/2\pi=$ 0 and 6.6 MHz), it is obvious that for the main diffraction ($0$th) order of the grating, increasing the spin excitation number causes the modulation effect opposite to adjust the spin position $z_s$. This is because the effective coupling strength is proportional to $n_s$ but inversely proportional to $z_s^ {-\frac{3}{2}}$ ($\tilde{\Omega}_{\mathrm{eff}} \propto n_s, z_s^ {-\frac{3}{2}}$).

\subsection{Beyond the weak probing limit with $vdW$ interaction}

Subsequently, we discuss the modulation for the grating under the scenarios beyond the weak probing limit. Firstly, we examine the optical response of target atomic ensembles at various positions under different probing intensities as Sect.~\ref{Weak}, including the analysis of absorption and dispersion curves and the average excitation probabilities of Rydberg states. 

\begin{figure}
\includegraphics[width=0.48\textwidth]{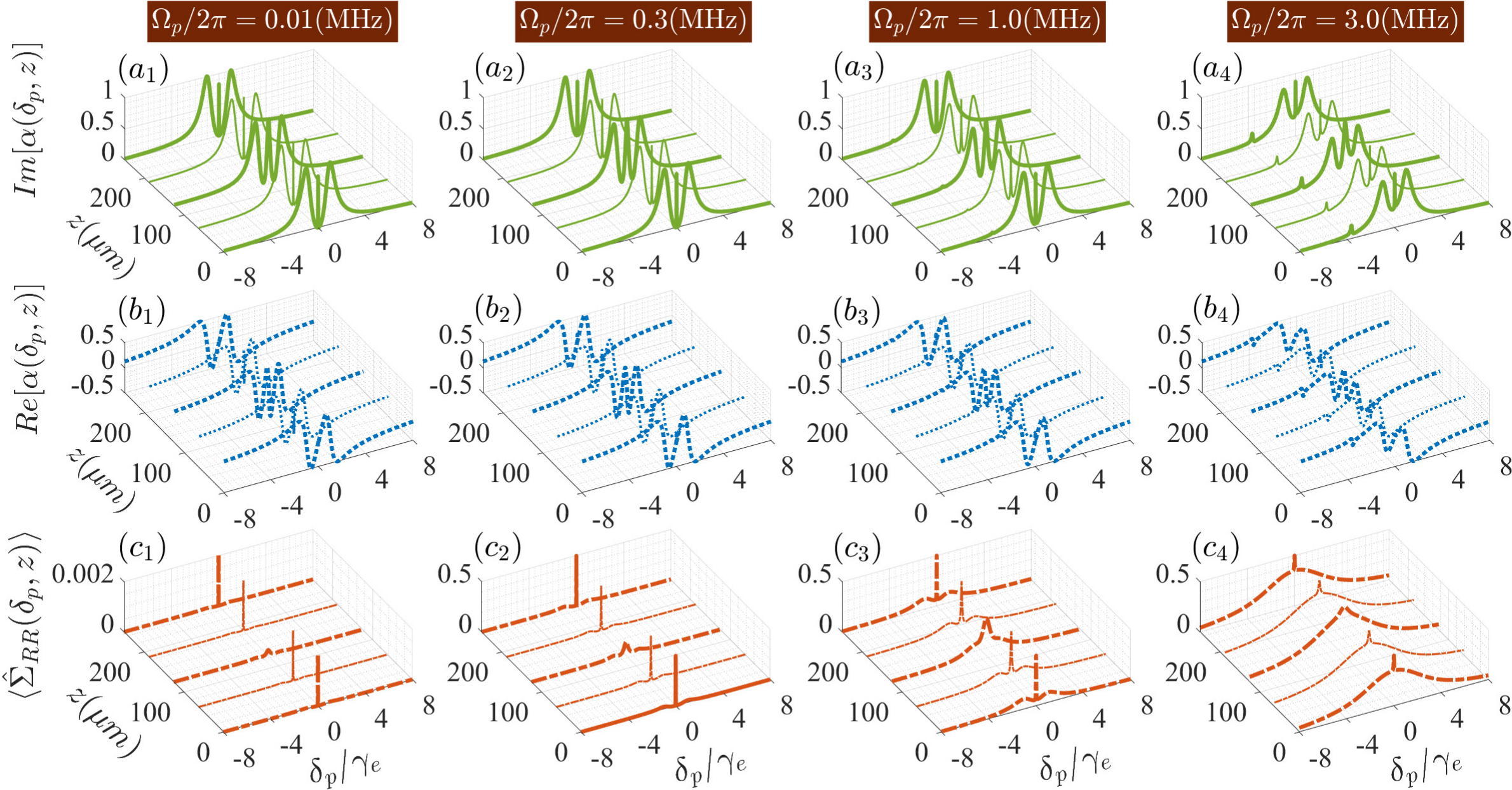}
\caption{Absorption $\langle\textbf{Im}[\hat{\alpha}]\rangle$ (olive-solid curves), dispersion $\langle\textbf{Re}[\hat{\alpha}]\rangle$ blue-dotted curves) and Rydberg population $\langle\hat{\Sigma}_{RR}\rangle$ (red-dashed and dotted curves) versus detuning $\delta_p$ and atomic position $z$ in the target ensemble for varying input probe intensity with the gate spin located at $z_s=L/2$, where, $n_s=500$ in panel ($a_1$). Other parameters are the same as those in Fig~\ref{Fig2}, except for $\Omega_p$. }
\label{Fig6}
\end{figure}

In Figure~\ref{Fig6}, we show absorption $\langle\textbf{Im}[\hat{\alpha}]\rangle$ (olive-solid), dispersion $\langle\textbf{Re}[\hat{\alpha}]\rangle$ (blue-dotted) and average Rydberg population $\langle\hat{\Sigma}_{RR}\rangle$ (orange-dashed and dotted) versus detuning $\delta_p$ and position $z$ in the target ensemble for different probe intensities, with the gate spin located at $z_s=L/2$. It is evident that with weak probing field $\Omega_p/2\pi=0.01$ MHz, the average excitation probability of Rydberg states becomes negligible $\langle\hat{\Sigma}_{RR}\rangle<0.002$ [See Fig.~\ref{Fig6}(c$_1$)], and the optical response of the atomic ensemble reverts to the scenario described in Sect.~\ref{Weak}. Based on the SA model \cite{David2011, Liu2014}, the average excitation probability of state $|r\rangle$ $\langle\hat{\Sigma}_{RR}\rangle$ increases with the rising of the probe intensity $\langle\hat{\mathcal{E}}^{\dag}_p(0)\hat{\mathcal{E}}_p(0)\rangle$, resulting in the emergence of blockade and cooperative effects. This is specifically manifested as the system's conditional polarizability being a superposition of a quasi-$\mathcal{N}$-type four-level system modulated by the spin position and an asymmetric $\Lambda$-type three-level system, unaffected by spin position but exhibiting ac Stark shift $\Delta=\Omega_d^2/\delta_d$, as shown in Eq.~\eqref{Polar}. 

\begin{figure}
     \includegraphics[width=0.48\textwidth]{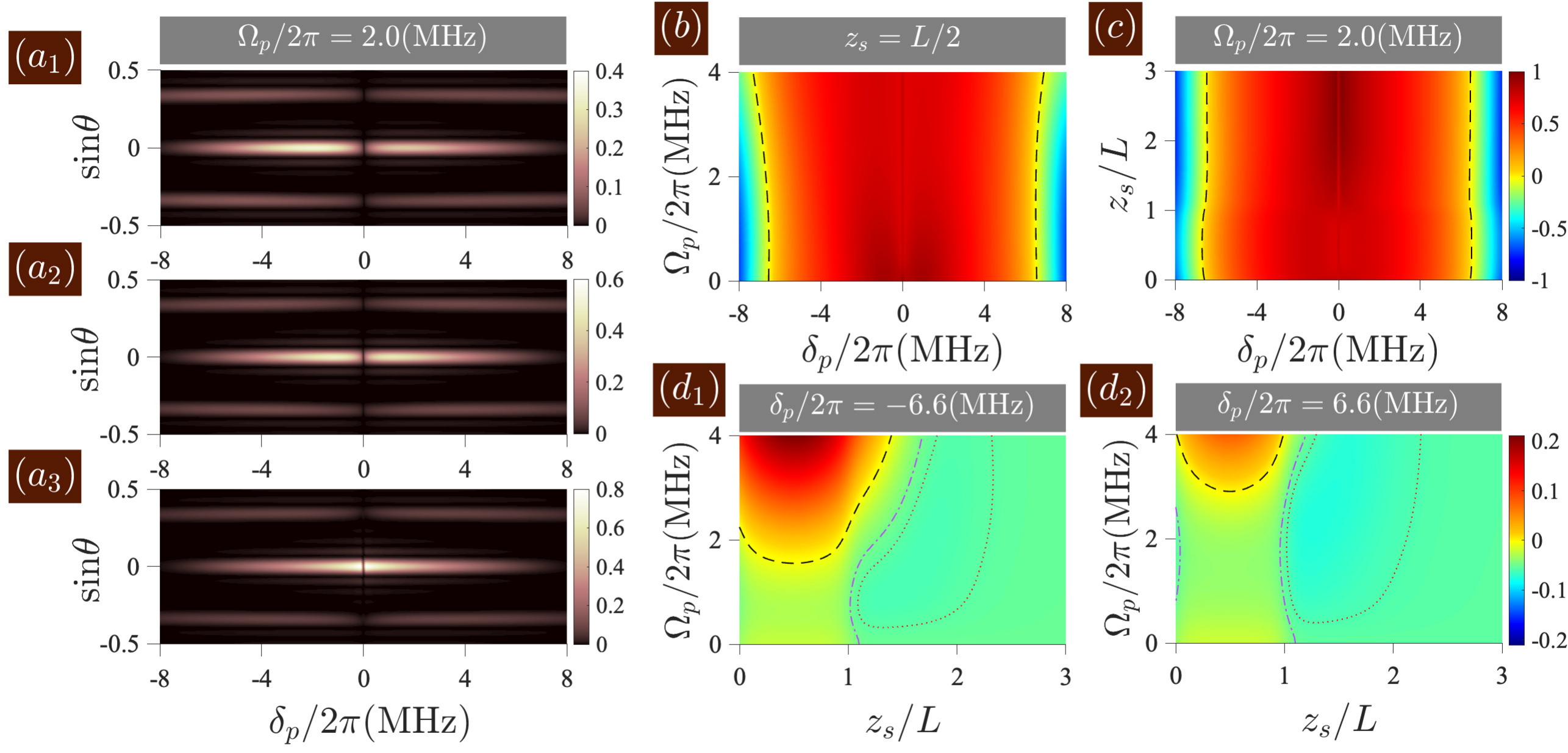}
     \caption{Fraunhofer diffraction intensity $I_{p}(\theta)$ versus detuning $\delta_{p} / 2 \pi$ and $\sin\theta$ with $\Omega_p/2\pi=2.0$ MHz for different spin locations $z_s=L/2, L$ and $3L$, corresponding panel (a$_1$)-(a$_3$). Relative diffraction intensity coefficient $\eta=\frac{I_p(\theta_0)-I_p(\theta_{\pm 1})}{I_p(\theta_0)+I_p(\theta_{\pm 1})}$ versus two of $\delta_p$, $\Omega_p$ and $z_s$ shown in right four panels (b)-(d$_2$). Specifically, plots are shown for $z_s=L/2$ in (b), $\Omega_p/2\pi=2.0$ MHz in (c), and $\delta_p/2\pi=\pm 6.6$ MHz in panels (d$_1$) and (d$_2$). And other parameters are the same as those in Fig.~\ref{Fig4}.}
     \label{Fig7}
     \end{figure}

 Then, we show in Fig.~\ref{Fig7}(a$_1$)-(a$_3$) the angular diffraction spectra versus the probe detuning $\delta_p$ for various spin positions $z_s$ under a strong input probe field $\Omega_p=2.0\times2\pi$ MHz. As we can see, in the frequency domain, diffraction intensity exhibits asymmetry at the spin position $z_s=L/2$, e.g. $I_p(\theta_0)=0.35$ at $\delta_p/2\pi=-2.0$MHz but $I_p(\theta_0)=0.2$ at $\delta_p/2\pi=2.0$MHz, which lessens as the spin moves away from the target atomic ensemble ($z_s \to 3L$). It is attributed to the heightened probability of Rydberg state excitation near the atomic ensemble [See Fig.~\ref{Fig6}], enhancing the contribution of the asymmetric three-level component (with ac Stark shift $\Delta=\Omega_d^2/\delta_d$) to the medium's effective susceptibility. 
 
 Aiming to further analyze the grating's diffraction characteristics, we then introduce a relative diffraction intensity coefficient, $\eta$, defined as $\eta=\frac{I_p(\theta_0)-I_p(\theta_{\pm 1})}{I_p(\theta_0)+I_p(\theta_{\pm 1})}$, ranging between $[-1,1]$. This coefficient reflects the tendency of the grating to diffract light primarily into the $\pm 1^{st}$ orders ($\eta \rightarrow -1$) or exhibit minimal diffraction ($\eta \rightarrow 1$). Figure~\ref{Fig7} illustrates the dependence of $\eta$ on the spin position $z_s$ and probe field intensity. Notably, Fig.~\ref{Fig7}~(d$_1$) and (d$_2$)  demonstrate the asymmetry in diffraction intensity at increased probe intensities for $\delta_p/2\pi=\pm 6.6$ MHz. The black-dashed lines represent $\eta=0$, indicating \textit{Damman-like} diffraction where $I_p(\theta_0)=I_p(\theta_{\pm 1})$. The magenta-dashed and red-dotted curves with $\eta=-0.05$ and $\eta=-0.06$, respectively, show that the grating's $\pm 1^{st}$ order normalized diffraction efficiency $I^N_p(\theta_{\pm 1})=I_p(\theta_{\pm 1})/\sum_i I_p(\theta_{i})$ reaches approximately 34.4\% and 34.6\%, a result of the combined modulation by spin position and probing field intensity.
 
\section{CONCLUSIONS}\label{SecIV}

In summary, spin-exchange-induced atomic gratings offer a captivating avenue, harnessing the spatial position ($z_s$) of spin atoms relative to the target atomic ensemble as novel degrees of freedom for non-local manipulation of grating diffraction properties. Our investigation of the far-field Fraunhofer diffraction properties of such new EIG reveals striking similarities to cooperative optical nonlinearity based on the Rydberg blockade effect. The optical response of the grating behaves as a superposition of $\Lambda$-type three-level and quasi-$\mathcal{N}$-type four-level atomic gratings. Remarkably, this scheme dispenses with nonlinear absorption, merely employing distance as a manipulation degree of freedom, and achieves similar diffraction properties avoiding reliance on the properties of the probe field. 

Furthermore, we find the number of spin excitations does impact the effective coupling strength but does not significantly modify the higher-order diffraction efficiency of the grating. Additionally, further discussing the impact of input probe intensity on the diffraction properties of the grating, it is observed that an increasing probe intensity induces asymmetry in the diffraction spectrum of the grating in the frequency domain. This work represents a significant contribution to the enrichment of manipulation degrees of freedom for gratings, thereby facilitating advancements in the nonlocal control of photon transport. The potential demonstrated by spin-exchange-induced atomic gratings opens up exciting possibilities for further exploration and utilization in optical manipulation research.

\section*{Acknowledgment}
This work is supported by the Natural Science Foundation of Jilin Province (20220101009JC); National Natural Science Foundation of China (12104107); Fundamental Research Funds for the Central Universities (2412022ZD046); D.-X. Li is supported by the Research Start-up Funds of the Shenyang Normal University (Grants No. BS202301, No. BS202303, No. BS202314).

\appendix

\section{Rydberg Excitation of Superatom}\label{SS}
\setcounter{equation}{0}
\renewcommand{\theequation}{A\arabic{equation}}
It is obvious that $|G\rangle$ and first-order collective states ($\left\vert E^{(1)}\right\rangle$, $\left\vert
M^{(1)}\right\rangle$ and $\left\vert R^{(1)}\right\rangle$) could form a four-level $\mathcal{N}$
configuration of superatom, neglecting other higher-order collective states safely in the weak probe limit. At this time, the Rydberg excitation probability in an SA is similar to that of a single atom. For a single atom with $\mathcal{N}$ configuration, we have the Heisenberg Langevin equation:
\begin{align}
\partial_{t}\hat{\varrho}_{gg} &  =2\gamma_{e}\hat{\varrho}_{ee}-i\hat{\Omega}_{p}\hat{\varrho}_{eg}+i\hat{\Omega}_{p}^{\dagger}\hat{\varrho}_{ge},\nonumber\\
\partial_{t}\hat{\varrho}_{mm} &  =2\gamma_{e}\hat{\varrho}_{ee}+2\gamma_{r}\hat{\varrho}_{rr}-i[\Omega_{c}\hat{\varrho}_{em}-\Omega_{c}^{\ast}\hat{\varrho}_{em}]\nonumber\\
& +i[\Omega_{\mathrm{eff}}J_{-}\hat{\varrho}_{mr}-\Omega_{\mathrm{eff}}^{*}J_{+} \hat{\varrho}_{r m} ],\nonumber\\
\partial_{t}\hat{\varrho}_{rr} &  =-2\gamma_{r}\hat{\varrho}_{rr}-i[\Omega_{\mathrm{eff}}J_{-}\hat{\varrho}_{mr}-\Omega_{\mathrm{eff}}^{*}J_{+} \hat{\varrho}_{r m} ],\nonumber\\
\partial_{t}\hat{\varrho}_{gm} &  =-[\gamma_m-i\Delta_{1}]\hat{\varrho}_{gm}+i\Omega_{\mathrm{eff}}J_{-} \hat{\varrho}_{gr}+i\Omega_{c}^{\ast}\hat{\varrho}_{ge},\nonumber\\
\partial_{t}\hat{\varrho}_{ge} &  =-[\gamma_{e}-i\Delta_{p}]\hat{\varrho
}_{ge}+i\hat{\Omega}_{p}\left[  \hat{\varrho}%
_{gg}-\hat{\varrho}_{ee}\right]+i\Omega_{c}^{*}\hat{\varrho}_{gm},\nonumber\\
\partial_{t}\hat{\varrho}_{gr} &  =-\left[  \gamma_{r}-i\left(  \Delta
_{p}+\Delta_{d}\right)  \right]  \hat{\varrho}_{gr}+i\Omega_{\mathrm{eff}}^{*}J_{+} \hat{\varrho}_{gm},\nonumber\\
\partial_{t}\hat{\varrho}_{me} &  =-[\gamma_{e}-i\Delta_{c}]\hat{\varrho
}_{me}+i\hat{\Omega}_{p}\hat{\varrho}_{mg}%
-i\Omega_{\mathrm{eff}}^{*}J_{+} \hat{\varrho}_{r e} \nonumber\\
&+i\Omega_{c}\left[  \hat{\varrho}_{mm}-\hat{\varrho}_{ee}\right]
,\nonumber\\
\partial_{t}\hat{\varrho}_{mr} &  =-\left[  \gamma_{r}-i\left(  \Delta
_{c}+\Delta_{d}\right)  \right]  \hat{\varrho}_{mr}-i\Omega_{c}%
\hat{\varrho}_{er}\nonumber\\
&+i\Omega_{\mathrm{eff}}^{*}J_{+} [\hat{\varrho}_{m m}-\hat{\varrho}_{r r}],\nonumber\\
\partial_{t}\hat{\varrho}_{er} &  =-[\gamma_{e}+\gamma_{r}-i\Delta
_{d}]\hat{\varrho}_{er}-i\hat{\Omega}_{p}^{\dagger}%
\hat{\varrho}_{gr}-i\Omega_{c}^{\ast}\hat{\varrho}_{mr}\nonumber\\
&+i\Omega_{\mathrm{eff}}^{*}J_{+}\hat{\varrho}_{em}.%
\label{A1}
\end{align}
Note also that, in an SA, the probe (coupling) coefficients are $\sqrt{n_{sa}}\hat{\Omega}_p$ ($\Omega_c$, $\Omega_{\mathrm{eff}}$) on transition $|G\rangle \to |E^{(1)}\rangle$ ($|E^{(1)}\rangle \to |M^{(1)}\rangle$, $|M^{(1)}\rangle \to |R^{(1)}\rangle$); $\sqrt{2(n_{sa}-1)}\hat{\Omega}_p$ ($\sqrt{2}\Omega_c$, $\sqrt{2}\Omega_{\mathrm{eff}}$) on transition $|E^{(1)}\rangle \to |E^{(2)}\rangle$ ($|E^{(2)}\rangle \to |E^{(1)}M^{(1)}\rangle$, $|E^{(1)}M^{(1)}\rangle \to |E^{(1)}R^{(1)}\rangle$); $\sqrt{n_{sa}-1}\hat{\Omega}_p$ ($\sqrt{2}\Omega_c$, $\Omega_{\mathrm{eff}}$) on transition $|M^{(1)}\rangle \to |E^{(1)}M^{(1)}\rangle$ ($|E^{(1)}M^{(1)}\rangle \to |M^{(2)}\rangle$, $|E^{(1)}M^{(1)}\rangle \to |E^{(1)}R^{(1)}\rangle$); etc. In this case, we obtain $\langle\hat{\Sigma}_{RR}(z)\rangle=\langle\hat{\Sigma}^{(1)}_{RR}(z)\rangle+\langle\hat{\Sigma}^{(2)}_{RR}(z)\rangle+\langle\hat{\Sigma}^{(3)}_{RR}(z)\rangle+...$, with the first-, second-, and third-order components.~As we notice from Eq.~(\ref{A1}) that
dynamic evolutions of these operators but with the replacement of $\hat{\Omega}_{p}^{\dag}(z)\hat{\Omega}_{p}(z)\to\eta_p\hat{\Omega}_{p}^{\dag}(z)\hat{\Omega}_{p}(z)$, $|\Omega_{c}|^2\to\eta_c|\Omega_{c}|^2$ and $|\Omega_{\mathrm{eff}}|^2\to\eta_d|\Omega_{\mathrm{eff}}|^2$ with $\eta_p$, $\eta_c$, $\eta_d$ being the corresponding substitution coefficients.~Thus, compared to the single atomic counterpart $\langle\hat{\varrho}_{\kappa\kappa}(z)\rangle=\overline{\varrho}_{\kappa\kappa}[\hat{\Omega}_p^{\dag}(z)\hat{\Omega}_p(z),|\Omega_c|^2,|\Omega_{\mathrm{eff}}|^2]$, with $\kappa\in \{e,r,m\}$, each order steady-state populations of SA can be expressed as:
\begin{align}\label{A2}
\langle\hat{\Sigma}_{EE}^{(1)}\rangle & =\overline{\varrho}_{ee}[n_{sa}\hat{\Omega}_p^{\dag}\hat{\Omega}_p,|\Omega_c|^2,|\Omega_{\mathrm{eff}}|^2],\nonumber\\
\langle\hat{\Sigma}^{(1)}_{MM}\rangle & =\overline{\varrho}_{mm}[n_{sa}\hat{\Omega}_p^{\dag}\hat{\Omega}_p,|\Omega_c|^2,|\Omega_{\mathrm{eff}}|^2],\\
\langle\hat{\Sigma}_{EE}^{(2)}\rangle & =\langle\hat{\Sigma}_{EE}^{(1)}\cdot\rangle\overline{\varrho}_{ee}[2(n_{sa}-1)\hat{\Omega}_p^{\dag}\hat{\Omega}_p,|\Omega_c|^2,|\Omega_{\mathrm{eff}}|^2],\nonumber\\
\langle\hat{\Sigma}^{(2)}_{MM}\rangle & =\langle\hat{\Sigma}^{(1)}_{MM}\rangle\cdot\overline{\varrho}_{mm}[(n_{sa}-1)\hat{\Omega}_p^{\dag}\hat{\Omega}_p,2|\Omega_c|^2,|\Omega_{\mathrm{eff}}|^2],\nonumber\\
\langle\hat{\Sigma}_{EE,MM}^{(1),(1)}\rangle & =\langle\hat{\Sigma}_{EE}^{(1)}\rangle\cdot\overline{\varrho}_{mm}[2(n_{sa}-1)\hat{\Omega}_p^{\dag}\hat{\Omega}_p,2|\Omega_c|^2,|\Omega_{\mathrm{eff}}|^2]\nonumber\\
&+\langle\hat{\Sigma}^{(1)}_{MM}\rangle\cdot\overline{\varrho}_{ee}[(n_{sa}-1)\hat{\Omega}_p^{\dag}\hat{\Omega}_p,|\Omega_c|^2,|\Omega_{\mathrm{eff}}|^2],\nonumber\\
\langle\hat{\Sigma}^{(1)}_{RR}\rangle & =\overline{\varrho}_{rr}[n_{sa}\hat{\Omega}_p^{\dag}\hat{\Omega}_p,|\Omega_c|^2,|\Omega_{\mathrm{eff}}|^2],\nonumber\\
\langle\hat{\Sigma}^{(2)}_{RR}\rangle & = \langle\hat{\Sigma}^{(1)}_{MM}\rangle\cdot \overline{\varrho}_{rr}[(n_{sa}-1)\hat{\Omega}^{\dag}_p\hat{\Omega}_p, 2|\Omega_c|^2, 2|\Omega_{\mathrm{eff}}|^2]\nonumber\\ & +\langle\hat{\Sigma}^{(1)}_{EE}\rangle \cdot \overline{\varrho}_{rr}[2(n_{sa}-1)\hat{\Omega}^{\dag}_p\hat{\Omega}_p, 2|\Omega_c|^2, |\Omega_{\mathrm{eff}}|^2],\nonumber\\
\langle\hat{\Sigma}^{(3)}_{RR}\rangle & = \langle\hat{\Sigma}^{(2)}_{EE}\rangle \cdot \overline{\varrho}_{rr}[3(n_{sa}-2)\hat{\Omega}^{\dag}_p\hat{\Omega}_p, 2|\Omega_c|^2, |\Omega_{\mathrm{eff}}|^2] \nonumber\\
 & +\langle\hat{\Sigma}_{EE,MM}^{(1),(1)}\rangle\cdot \overline{\varrho}_{rr}[2(n_{sa}-2)\hat{\Omega}^{\dag}_p\hat{\Omega}_p, |\Omega_c|^2, 2|\Omega_{\mathrm{eff}}|^2]\nonumber\\
 & +\langle\hat{\Sigma}^{(2)}_{MM}\rangle \cdot \overline{\varrho}_{rr}[(n_{sa}-2)\hat{\Omega}^{\dag}_p\hat{\Omega}_p, 3|\Omega_c|^2, 3|\Omega_{\mathrm{eff}}|^2].\nonumber
\end{align}

We write down the collective states of SA,

\begin{align}\label{A3}
|E^{(j)}\rangle & =\frac{[\sum_{i=1}^{n_{sa}}|e_i\rangle\langle g_i|e^{ik_pz_i}]^j}{\sqrt{n_{sa}!j!/(n_{sa}-j)!}}|G\rangle,\\
|M^{(j)}\rangle & =\frac{[\sum_{i=1}^{n_{sa}}|m_i\rangle\langle g_i|e^{i(k_p-k_c)z_i}]^j}{\sqrt{n_{sa}!j!/(n_{sa}-j)!}}|G\rangle,\nonumber\\
|R^{(j)}\rangle & =\frac{[\sum_{i=1}^{n_{sa}}|r_i\rangle\langle g_i|e^{i(k_p-k_c+k_r)z_i}]^j}{\sqrt{n_{sa}!j!/(n_{sa}-j)!}}|G\rangle,\nonumber\\
 |E^{(s)}M^{(j)}R^{(1)}\rangle & =\frac{[\sum\limits^{n_{sa}}_{\mu=1}|e_{\mu}\rangle\langle g_{\mu}|e^{ik_pz_{\mu}}\cdot e^{ik_p z_{\mu}}]^s}{\sqrt{n_{sa}!s!j!/(n_{sa}-s-j-1)!}}\nonumber\\
  & \cdot [\sum\limits^{n_{sa}}_{\mu=1}|m_{\mu}\rangle\langle g_{\mu}|\cdot e^{i(k_p-k_c) z_{\mu}}]^j\nonumber\\
  & \cdot \sum\limits^{n_{sa}}_{\mu=1}|r_{\mu}\rangle\langle g_{\mu}|\cdot e^{i(k_p-k_c+k_r) z_{\mu}}\left\vert G\right\rangle.\nonumber
 \end{align}

\section{Quasi-one-dimensional approximation}\label{Omega}
\setcounter{equation}{0}
\renewcommand{\theequation}{A\arabic{equation}}
\setcounter{figure}{0}
\renewcommand{\thefigure}{A\arabic{figure}}

A quasi-one-dimensional approximation has been utilized to analyze the diffraction properties of gratings in our scheme. However, the dipole-exchange interactions are not limited to the $z$-axis alone. It is necessary to examine the potential impacts of deviations introduced by orthogonal directions, e.g., $y$-direction, on the diffraction behavior of the grating.

\begin{figure}\includegraphics[width=0.48\textwidth]{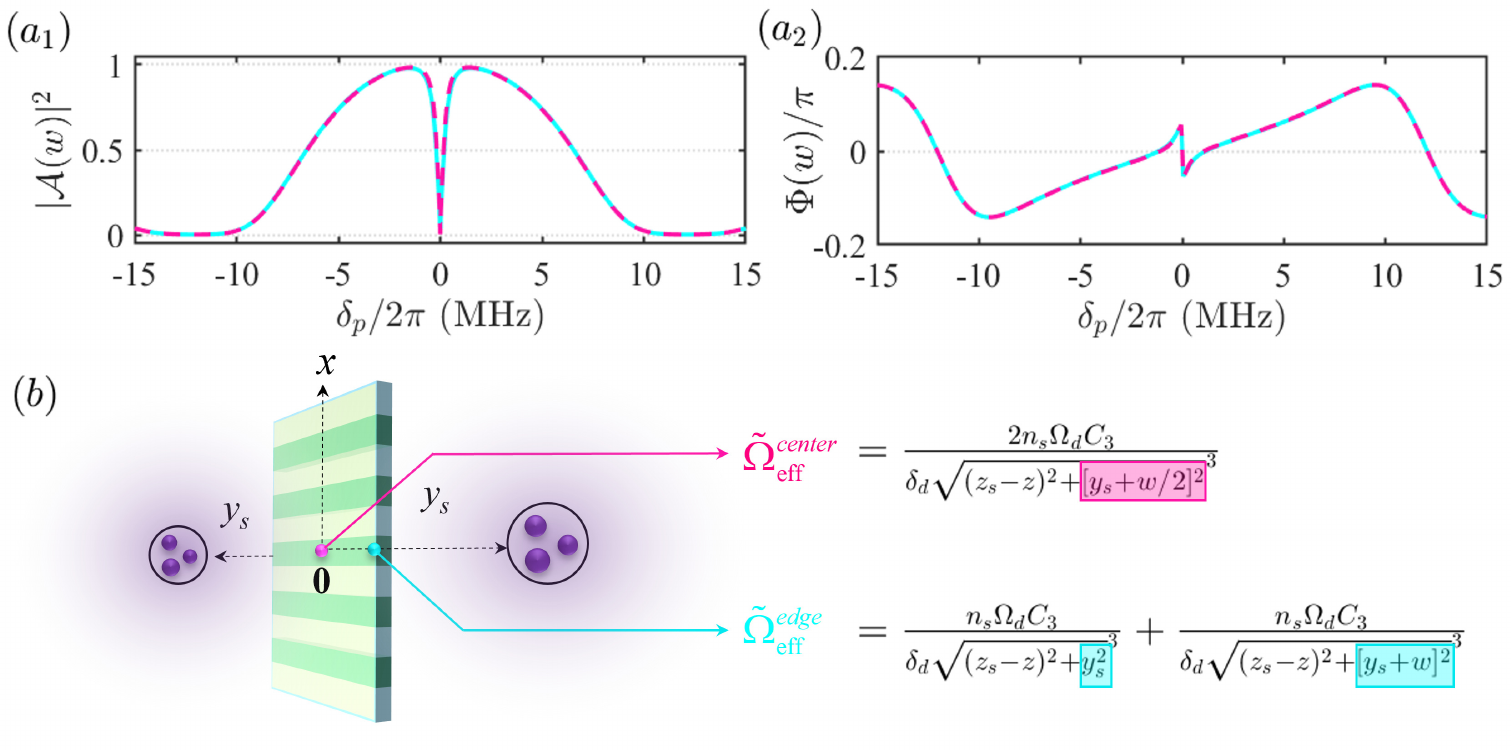}
     \caption{(a$_1$) Amplitude transmission rate $|\mathcal{A}(\omega)|^2$ and (a$_2$) phase $\phi_{p}(L) / \pi$ versus detuning $\delta_{p} / 2 \pi$ with the effective coupling $\Omega^{c}_{\mathrm{eff}}$ (magenta-dashed curves at the center) and $\Omega^{e}_{\mathrm{eff}}$ (aqua-solid curves at the edge). (b) Schematic of the two positions of the grating medium with max error in the $y$-direction and their effective coupling. Here we set parameters the same as in Fig.~\ref{Fig2}.}
     \label{Fig8}
     \end{figure}

In the above calculations, we employed a scheme where gate atoms' spins were symmetrically placed on both sides of the atomic ensemble in the $y$ direction. Throughout the calculations, we assumed that the potential experienced by atoms in the $y$ direction is uniform, implying that the effective coupling strength is also uniform and denoted as $\tilde{\Omega}_{\mathrm{eff}}\simeq\tilde{\Omega}_{\mathrm{eff}}^c=\frac{2n_s\Omega_d C_3}{\delta_d\sqrt{(z_s-z)^2+[y_s+w/2]^2}^3}$. However, at any arbitrary position in the $y$ direction, the effective coupling strength is $\tilde{\Omega}_{\mathrm{eff}}=\frac{n_s\Omega_d C_3}{\delta_d\sqrt{(z_s-z)^2+y^2}^3}+\frac{n_s\Omega_d C_3}{\delta_d\sqrt{(z_s-z)^2+[2y_s+w-y]^2}^3}$, as illustrated in Fig.~\ref{Fig8}, with $w$ representing the width of the target atomic ensemble along the $y$ axis, with $w=Ma=5a\simeq8\mu$m. Fig.~\ref{Fig8}($a_1$) and ($a_2$) respectively present the amplitude transmission $|\mathcal{A}(\omega)|^2$ and phase $\phi(\omega)$ spectra of the target atomic ensemble calculated using two different computational methods, with the magenta-dashed curves and aqua-solid curves corresponding to effective coupling strengths $\tilde{\Omega}_{\mathrm{eff}}^c=\frac{2n_s\Omega_d C_3}{\delta_d\sqrt{(z_s-z)^2+[y_s+w/2]^2}^3}$ and $\tilde{\Omega}_{\mathrm{eff}}^e=\frac{n_s\Omega_d C_3}{\delta_d\sqrt{(z_s-z)^2+y_s^2}^3}+\frac{n_s\Omega_d C_3}{\delta_d\sqrt{(z_s-z)^2+[y_s+w]^2}^3}$. Here $\tilde{\Omega}_{\mathrm{eff}}^c$ and $\tilde{\Omega}_{\mathrm{eff}}^e$ represent the equivalent Rabi frequencies of the medium at the center and edges in the $y$-direction, respectively. It is easy to find that the optical responses from the two positions with the greatest differences in $y$-direction are almost identical, in Fig.~\ref{Fig8} (a$_1$) and (a$_2$), indicating that the errors in the y-direction can be safely ignored, thereby confirming the effectiveness of our approximation method.

\bibliographystyle{apsrev4-1}
\bibliography{Manu.bib}
\end{document}